\documentclass{elsart}

\usepackage{epsfig}

\usepackage{amssymb}

\begin{document}

\begin{frontmatter}



\title{Drift chamber with a c-shaped frame}


\author[cracow]{J. Smyrski\corauthref{corr}},
\ead{smyrski@if.uj.edu.pl}
\author[julich1]{Ch. Kolf},
\author[munster]{H.-H. Adam},
\author[bronowic]{A. Budzanowski},
\author[cracow]{R. Czy\.zykiewicz},
\author[julich1]{D. Grzonka},
\author[cracow]{A. Heczko},
\author[cracow]{M. Janusz},
\author[cracow]{L. Jarczyk},
\author[cracow]{B. Kamys},
\author[munster]{A. Khoukaz},
\author[julich1]{K. Kilian},
\author[katowice]{P. Kowina},
\author[cracow]{A. Misiak},
\author[cracow]{P. Moskal},
\author[julich1]{W. Oelert},
\author[cracow]{C. Piskor-Ignatowicz},
\author[cracow]{J. Przerwa},
\author[munster]{C. Quentmeier},
\author[katowice]{T. Ro\.zek},
\author[munster]{R. Santo},
\author[julich1]{G. Schepers},
\author[julich1]{T. Sefzick},
\author[katowice]{M. Siemaszko},
\author[munster]{A. T\"aschner},
\author[julich1]{P. Winter},
\author[julich1]{M. Wolke},
\author[julich2]{P. W\"ustner},
\author[katowice]{W. Zipper}

\address[cracow]{Institute of Physics, Jagellonian University, Pl-30-059 Cracow, Poland}
\address[julich1]{IKP, Forschungszentrum J\"ulich, D-52425 J\"ulich, Germany}
\address[munster]{Institut f\"ur Kernphysik, Westf\"alische Wilhelms-Universit\"at, 
D-48149 M\"unster, Germany}
\address[bronowic]{Institute of Nuclear Physics, Pl-31-342 Cracow, Poland}
\address[katowice]{Institute of Physics, University of Silesia, PL-40-007 Katowice, Poland}
\address[julich2]{ZEL, Forschungszentrum J\"ulich, D-52425 J\"ulich, Germany}

\corauth[corr]{Corresponding author. Correspondence address: Institute of Physics, 
Jagellonian University, ul. Reymonta 4, Pl-30-059 Cracow, Poland. Tel.:+48-12-663-5616; 
fax: +48-12-634-2038.}

\begin{abstract}
We present the construction of a planar drift chamber with wires stretched between
two arms of a c-shaped aluminium frame. 
The special shape of the frame allows to extend
the momentum acceptance of the COSY-11 detection system towards lower momenta 
without suppressing the high momentum particles.
The proposed design allows for construction of tracking detectors covering small angles 
with respect to the beam, which can be installed and removed without dismounting the beam-pipe. 
For a three-dimensional track reconstruction
a computer code was developed using a simple algorithm of hit preselection.

\end{abstract}

\end{frontmatter}

\section{Introduction}
\label{Introduction}

\begin{figure}[h]
\centering
\leavevmode                      
\epsfverbosetrue
\epsfclipon
\epsfxsize=8cm
\epsffile[50 300 350 550]{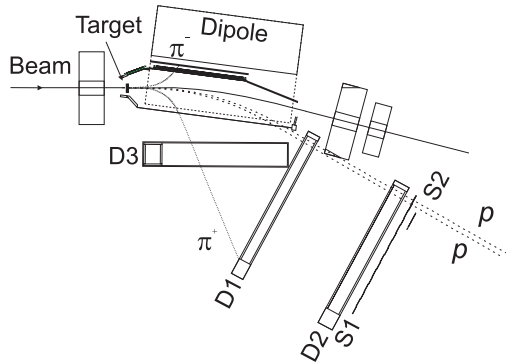}
\caption{COSY-11 detection system. Indicated particle tracks correspond to
a $pp \rightarrow pp \pi^+ \pi^-$ event simulated for the beam momentum
exceeding by 1~MeV/c the threshold momentum equal to 1219~MeV/c. 
The two outgoing protons are registered with a set of two chambers --- D1 and D2
and the $\pi^+$ trajectory is measured with the presently introduced chamber --- D3.
The scintillation hodoscopes S1 and S2 are used as start detectors 
and S3 hodoscope placed in a distance of 9.4~m from S1 and not indicated in the figure 
is used as stop detector for time of flight measurements.}
\label{COSY-11}
\end{figure}

The drift chamber which is described in this report was built for
the COSY-11 experimental facility operating at the Cooler Synchrotron COSY-J\"ulich~\cite{Ma}.
The COSY-11, described in details in Ref.\ \cite{Br} and shown schematically 
in Fig.\ \ref{COSY-11}, is a magnetic spectrometer  
for measurements at small angles and is
dedicated to studies of near-threshold meson production in proton--proton
and proton--deuteron collisions (see e.g. Refs.\ \cite{PM,Qu,Mo}).
It uses one of the regular COSY dipole magnets for momentum analysis of charged reaction products
originating from interaction of the internal COSY beam particles 
with the nuclei of a cluster beam target~\cite{Do}.
Trajectories of positively-charged particles
which are deflected in the dipole magnet
towards the center of the COSY-ring are registered with a set of two planar drift chambers
indicated as D1 and D2 in Fig.\ \ref{COSY-11}.
These chambers cover only the upper range of momenta of the outgoing particles
what suffices to measure e.g. two outgoing protons in 
the $pp \rightarrow pp\eta'$ process~\cite{Mo}
or $p-p-K^+$ tracks in the $pp \rightarrow ppK^+K^-$ reaction 
close to threshold~\cite{Qu}.
For tracking positively charged pions appearing in near-threshold reactions
such as $pp \rightarrow pp \pi^{+}\pi^{-}$ with momenta by a factor of $m_{\pi}/m_{p}$ smaller
then the proton momenta it was necessary to extend the COSY-11 momentum acceptance towards smaller
values. 

Another important purpose was the detection of positively charged kaons prior to their decay
what is especially important for the measurement of the $pp \rightarrow ppK^+K^-$ 
close to threshold due to its small cross section on the level of a few nanobarns.
For this, an additional drift chamber was built and installed in the free space 
along the COSY-11 dipole magnet (see chamber D3 in Fig.\ \ref{COSY-11}) .
The main requirement for the chamber was 
a shape of the supporting frame which would not interfere with high momentum particles
that are registered in the detectors D1 and D2.
This demand was fulfilled by choosing  the frame of a rectangular form
but with one vertical side missing, called c-shaped frame since 
it resembles the character c.
The main design characteristics of the chamber are given in section 2. 
It was also essential, that the chamber allows for a three-dimensional track reconstruction
in order to determine the momentum vectors of registered particles 
at the target by tracing back their trajectories in the magnetic field 
of the COSY-11 dipole magnet to the nominal target position.
Therefore three different wire orientations were chosen and a track reconstruction
program was developed. This program is described in section 3.
The chamber calibration and results of the track reconstruction are presented in section 4.

\section{The chamber construction}
\label{Chamber}

\begin{figure}[ht]
\centering
\leavevmode                      
\epsfverbosetrue
\epsfclipon
\epsfxsize=5.5cm
\epsffile[180 580 350 760]{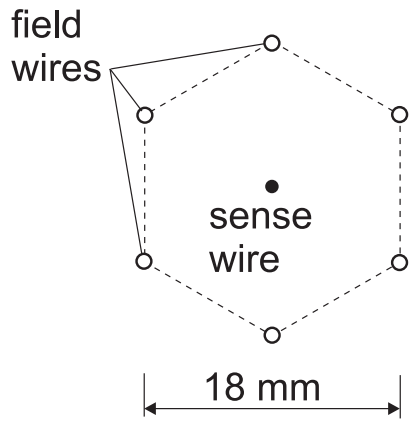}
\caption{Hexagonal cell.}
\label{cell}
\end{figure}

The sensitive chamber volume is built up by hexagonal drift cells (see Fig.\ \ref{cell}) 
identical with the structure used in the central drift chamber of the SAPHIR detector~\cite{Sc}.
In this type of cells the drift field has approximately cylindrical
symmetry, and thus the distance-drift time relation 
depends only weekly on the particles' angle of incidence.
In order to minimize the multiple scattering on wires, 
gold-plated aluminium~\cite{Al} is used for the 110~$\mu$m-thick field wires,
whereas the sense wires are made of 20~$\mu$m-thick gold-plated tungsten.

\begin{figure}[ht]
\centering
\leavevmode
\epsfverbosetrue
\epsfclipon
\epsfxsize=7.5cm
\epsffile[110 460 440 630]{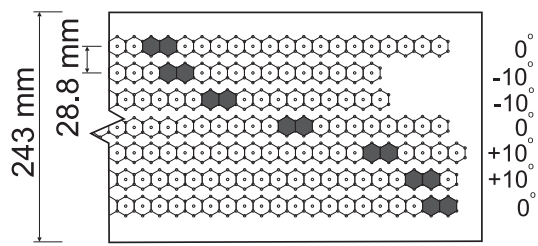}
\caption{Upper end-plate with indicated positions of openings for feedthrough 
used for mounting the wires.}
\label{plate}
\end{figure}

The cells are arranged in seven detection planes as indicated in Fig.\ \ref{plate}
showing one of two parallel aluminium endplates between which the wires are stretched.
Three detection planes (1, 4 and 7) contain vertical wires,
two planes (2, 3) have wires inclined at $-10^{\circ}$ and the remaining two planes (5, 6) 
contain wires inclined at $+10^{\circ}$.
This arrangement makes it possible to reconstruct particle
trajectories in three dimensions, also in cases of multi-track events.
The relatively small inclination of the wires was chosen since the measurement
of particle trajectories in the horizontal plane needs to be much more accurate compared to the
vertical plane. This is due to the fact that the particle
momenta are determined on the basis of the deflection of their trajectories in an almost uniform vertical 
magnetic field in the 6~cm gap of the COSY-11 dipole magnet. 

Each  endplate contains bore holes of \O4~mm with inserted feedthroughs for mounting the wires.
The drilling was done with a CNC jig boring machine assuring a positioning accuracy
better than $\pm$0.01~mm.
The holes for the field wires form equilateral hexagons with a width of 18~mm. 
In the middle of these hexagons there are the openings for the sense wires. 
The resulting spacing of the sense wires in the planes with vertical wires 
is again equal to 18~mm and in the detection planes with wires inclined by $\pm 10^{\circ}$ 
is equal to 18~mm$\times \cos 10^{\circ}$ = 17.73~mm.
The chamber contains $3 \times 80 = 240$ cells with vertical wires, $2 \times 76 = 152$ cells 
with wires inclined by $-10^{\circ}$ and $2 \times 77 = 154$ cells with wires inclined by $+10^{\circ}$.

The disturbance of the electric field in the drift cells of a given plane
by the cells of the neighboring planes was investigated
using the Garfield code~\cite{Ga}.
With the chosen spacing of the detection planes equal to 28.8 mm (see Fig.\ \ref{plate})
the resulting corrections to the distance-drift time relation determined as a function
of position along the sense wires are smaller than 0.05~mm~\cite{Ko}
for distances from the sense wire smaller than half of the cell width (9~mm). 
These corrections were neglected in view of the expected precision of track reconstruction in the
order of 0.1~mm. Thus one can assume that the distance-drift time relation 
is the same for all the cells in a given detection plane, what simplifies the chamber calibration.

\begin{figure}[ht]
\centering
\leavevmode
\epsfverbosetrue
\epsfclipon
\epsfxsize=9cm
\epsffile[95 460 495 710]{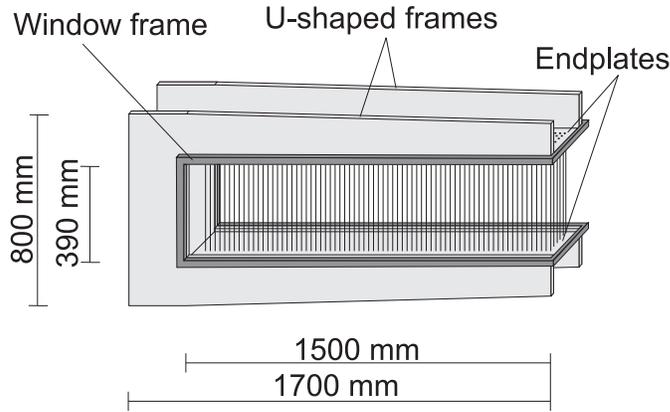}
\caption{Schematic three-dimensional view of the chamber frame for mounting the wires.
The window for particles is 1500~mm wide and 390~mm high.}
\label{view}
\end{figure}

The two aluminium endplates for mounting the wires are 15~mm thick
and are supported by two c-shaped frames made out of 20~mm thick aluminium 
(see a three-dimensional view in Fig.\ \ref{view}).
The frames hold the total load of about 2.4~kN 
originating from the mechanical tension of $\sim$0.2~kN ($= 546 \times 0.3$~N)  of the 
sense wires  and $\sim$2.2~kN ($= 2198 \times 1$~N) of the field wires.
Prior to mounting the wires, the frames were pre-stressed with a force corresponding
to the tension of wires using led bricks uniformly distributed
on the upper endplate.
This caused a deflection of the endplates at the free ends of 0.7~mm each.
The ends of the pre-stressed endplates  were fixed together using a steel plate and the bricks
were taken away. 
After mounting the wires the steel plate was removed and we checked that 
the distance between the ends of the endplates did not change.

\begin{figure}[ht]
\centering
\leavevmode
\epsfverbosetrue
\epsfclipon
\epsfxsize=7.5cm
\epsffile[90 390 500 655]{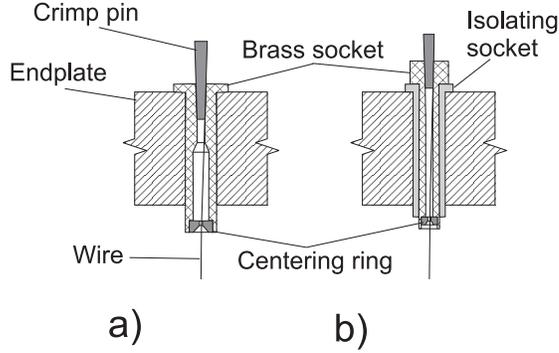}
\caption{Section of feedthrough for (a) field wires, and (b) sense wires.}
\label{feed}
\end{figure}

\begin{figure}[ht]
\centering
\leavevmode
\epsfverbosetrue
\epsfclipon
\epsfxsize=7.5cm
\epsffile[85 465 470 705]{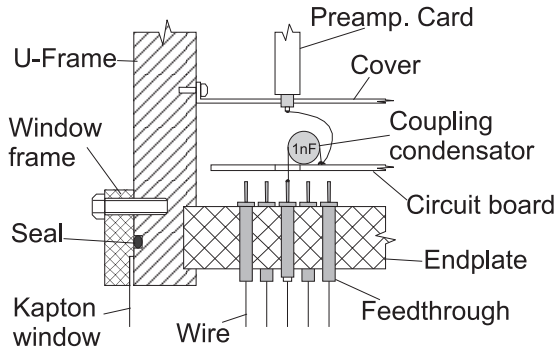}
\caption{Section of the upper endplate.}
\label{cross}
\end{figure}

For mounting the field wires brass feedthroughs with a \O150~$\mu$m inner opening are used.
Since all field wires should have the same ground potential 
electrical contact of the feedthroughs with the aluminium endplates
is assured by usage of a conductive glue.
For the sense wires we used isolating feedthroughs made of Delrin, a polyacetal
with a high dielectric strength of about 40~kV/mm, with inserted brass
feedthroughs containing \O50~$\mu$m openings for the positioning of the wires.
A schematic drawing of the feedthroughs is shown in Fig.\ \ref{feed}.
The wires were strung between pairs of feedthroughs and were fixed by means of copper crimp pins.
For sealing the feedthroughs and the crimp pins epoxy resin was used.
The high voltage was connected to the sense wires through 1~M$\Omega$ resistors
soldered directly to the crimp pins on the lower endplate. 
The signals were read out through 1~nF coupling capacitors connected to the sense wires
on the upper endplate (see Fig.\ \ref{cross}).
The capacitors and resistors were enclosed in hermetical volumes which were dried out
with silica aerogel allowing to reduce leakage currents.

For the window for particles penetrating the chamber 20~$\mu m$ thick Kapton foil is used 
glued to an aluminium frame which is screwed on the support frame and sealed with \O3~mm 
o-ring gasket.

Each of the detection planes is equipped with five 16-channel preamplifier-discriminator
cards \cite{Ja} based on the Fujitsu MB13468 amplifier chip 
and the LeCroy MVL407S comparator chip. 
The cards are mounted directly on the chamber.
The output pulses in the ECL standard are led by means of 30 m long
twisted-pair cables to FASTBUS-TDCs of LeCroy 1879 type
working in the common stop mode
with the pulses from the discriminators as start signals 
and the trigger pulse as the common stop.

As chamber gas, P10 mixture (90\% argon and 10\% methane) at atmospheric 
pressure is used.
The gas flow through the chamber is about 12~l/h.
The sense wire potential is +1800~V, whereas the field wires are all
grounded.
The gas amplification in the chambers is about $10^{5}$ and the discrimination threshold
set in the preamplifier-discriminator cards corresponds to $3 \cdot 10^{5}$ electrons.

\section{Track reconstruction algorithm}
\label{Track}

\begin{figure}[ht]
\centering
\leavevmode
\epsfverbosetrue
\epsfclipon
\epsfxsize=7.5cm
\epsffile[42 515 280 767]{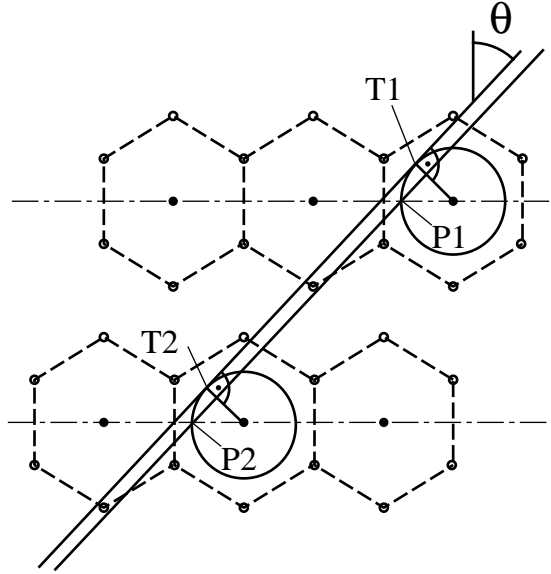}
\caption{Illustration of iterative procedure of fitting a straight line to two hits.}
\label{pair}
\end{figure}

For the reconstruction of particle tracks a simple algorithm was developed 
and implemented as a computer code written in the C-language. 
The reconstruction proceeds in three stages:
\begin{itemize}
\item
finding track candidates in two dimensions, independent for 
each orientation of the wires,
\item
matching the two-dimensional solutions in three dimensions,
\item
three-dimensional fitting in order to obtain optimal track parameters.
\end{itemize}
In the first stage, track candidates are found in two-dimensions 
using hits in the detection planes with the same inclination of the wires. 
For this, all possible combinations of pairs of hits from two different
detection planes are taken into account. 
Due to the left-right ambiguity of the track position with respect to the sense wire,
for each pair of hits there are four straight-line solutions possible. 
These solutions are determined using an iterative procedure.
First, the track distance to the sense wire is calculated 
from the drift time to drift distance relation  $d(t,\theta)$ 
assuming the track entrance angle $\theta$ as equal to $0^{\circ}$.
Note that the time-to-distance function may depend on the angle $\theta$ (see Fig.\ \ref{pair})
and this is taken into account when making a calibration.
Thus, the track is defined by a pair of points lying in the corresponding 
sense wire planes and indicated as P1 and P2 in Fig.\ \ref{pair}. 
In the next step, the inclination of this track is taken for determining  
new values of the distances from the sense wires and constructing a new pair of points defining 
the track. These points are indicated as T1 and T2 in Fig.\ \ref{pair}.
This procedure can be repeated, however, we terminate it 
already after the second step since further steps give negligible corrections. 
For each track determined by this method, one  subsequently finds hits 
which are consistent with it within a certain corridor along the track.
In this way, hits from
neighboring cells with respect to the selected two cells and from other detection planes with
the same orientation of wires are taken into account in the reconstruction.
For the width of the corridor we take $\pm$1~mm which is a few times larger than 
the position resolution of a single detection plane. With this choice more than 90\% of all
hits are included in the reconstruction as it is discussed in the next chapter. 

\begin{figure}[ht]
\centering
\leavevmode
\epsfverbosetrue
\epsfclipon
\epsfxsize=4cm
\epsffile[68 629 230 813]{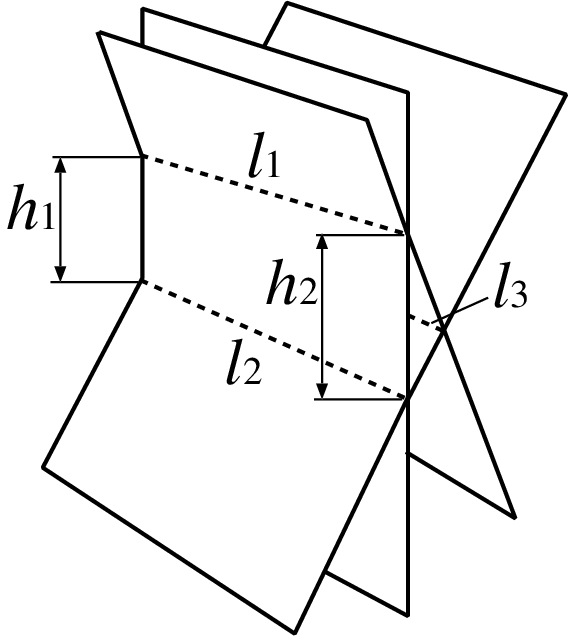}
\caption{Crossing of planes representing two-dimensional solutions 
for three different orientations of wires.}
\label{match}
\end{figure}

In three dimensions a two-dimensional solution can be represented by a plane containing 
the track and parallel to the corresponding sense wires.
Planes representing the two-dimensional solutions for three different
directions of sense wires used in the chamber 
should intersect along a common line corresponding to the particle track.
However, due to the limited spatial resolution of the chamber, the planes intersect along three
different lines (see Fig.\ \ref{match}).
Two lines --- $l_1$ and $l_2$ correspond to the intersections of the inclined planes 
with the vertical plane and the third line --- $l_3$ is an intersection of the inclined planes.
In order to quantify the  consistency between the two-dimensional solutions, 
we determine the distances $h_1$ and $h_2$ between the crossing points 
of the lines $l_1$, $l_2$ with the first and the seventh detection plane, respectively 
(see Fig.\ \ref{match}). 
For the three-dimensional reconstruction, we accept only the combinations of two-dimensional solutions
for which the distances $h_1$ and $h_2$ are both smaller then a certain limit $h$ 
which is adjusted as a reconstruction parameter. 
Too large values of $h$ lead to an unnecessary increase of CPU time 
since too many non-matching combinations are taken into account. 
On the contrary, too small values of $h$  cause losses of correct combinations 
and consequently decrease the reconstruction efficiency. In our case, a reasonable value
for $h$ is 2~cm.

\begin{figure}[ht]
\centering
\leavevmode
\epsfverbosetrue
\epsfclipon
\epsfxsize=6cm
\epsffile[17 613 260 835]{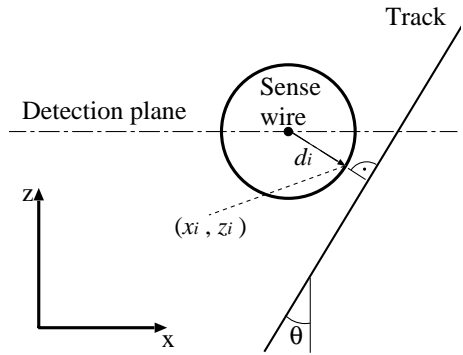}
\caption{Definition of the point $(x_i, z_i)$ used in the two-dimensional track reconstruction.}
\label{fit}
\end{figure}

The determination of track parameters in three dimensions proceeds via minimization 
of the $\chi^2$ function calculated on the basis of the differences between the measured 
distances of tracks from sense wires $d_i$ and the distances of the fitted tracks 
from the sense wires $d_i^f$:
\begin{equation}
\chi^2 = \sum_{i}\frac{(d_i-d_i^f)^2}{(\delta d_i)^2},
\end{equation}
where the summation proceeds over all hits assigned to one track 
and $\delta d_i$ is an uncertainty of the measured distance $d_i$.
In the case of the two-dimensional track reconstruction for wires oriented 
in one direction, the numerator in the above expression can be easily expressed 
in Cartesian coordinates using the formula:
\begin{equation}
(d_i-d_i^f)^2 = \frac{[x_i-(a z_i+b)]^2}{(1+a^2)},
\end{equation}
where $a$ and $b$ are parameters in the straight line equation: $x =a z + b$ representing the track,
$x$ and $z$ are coordinates  perpendicular to the sense wires and $(x_i, z_i)$ is a point
lying in the distance $d_i$ from the sense wire in the location corresponding
to the closest approach to the track (see Fig.\ \ref{fit}).
For the three-dimensional reconstruction, we use a coordinate system
with the $y$-axis parallel to the vertical wires and the z-axis perpendicular
to the detection planes. The origin of the coordinate system is located in 
the first plane.
Particle tracks are parametrized as straight lines in a conventional way:
\begin{eqnarray}
x' &=& a' z' + b',\\
y' &=& c' z' + d',
\end{eqnarray}
where $a',b' ,c' ,d'$ are the searched track parameters.
They can be linked to the parameters $a, b$ corresponding to the two-dimensional solutions 
using the linear transformation:
\begin{eqnarray}
a &=& a' \cos\alpha_i + c' \sin\alpha_i,\\
b &=& b' \cos\alpha_i + d' \sin\alpha_i,
\end{eqnarray}
where $\alpha_i$ is the angle between the sense wires and the vertical direction.
Inserting Eqs.\ (5) and (6) into (2) and the result into (1), the expression for $\chi^2$ takes the form:
\begin{equation}
\chi^2 = \sum_{i}\frac{[x_i-(a'\cos\alpha_i + c' \sin\alpha_i)z_i-
(b' \cos\alpha_i + d' \sin\alpha_i)]^2}
{[1+(a' \cos\alpha_i + c' \sin\alpha_i)^2](\delta d_i)^2}.
\end{equation}
The derivative of the dimensionless term in the square bracket in the denominator  
is of the order of unity and is negligible compared with corresponding derivative of the numerator 
divided by $(\delta d_i)^2$ which is of the order of $z_i / \delta d_i$.
Therefore, during the fitting procedure we regard the numerator just
as a constant  given by the initial values of the parameters 
and thus the minimization of $\chi^2$ is reduced to the linear least squares problem.
For the fitting we use the {\em svdfit} function from ``Numerical Recipes in C''~\cite{Nu}.
The three-dimensional track fitting is performed for all combinations 
of two-dimensional track candidates
and the solution with the minimal value of $\chi^2$ is selected. 
The hits corresponding to this solution are
then removed from the set of hits recorded in a given event 
and the reconstruction is repeated from the beginning
until there are no further candidates of tracks. 
In this way multi-track events can be reconstructed.

\section{Calibration and reconstruction results}
\label{Calibration}

\begin{figure}[ht]
\centering
\leavevmode
\epsfverbosetrue
\epsfclipon
\epsfxsize=6cm
\epsffile[0 0 500 480]{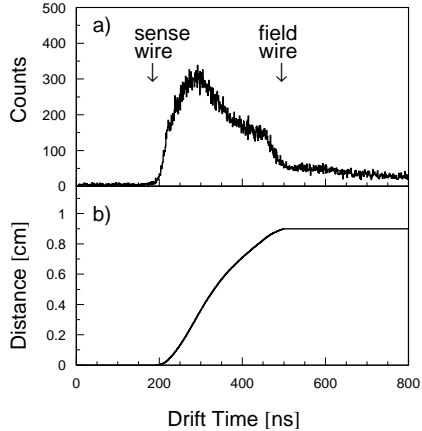}
\caption{Drift time spectrum for the first detection plane (a) and space-time relationship
obtained by integration of the drift time spectrum in the time interval indicated by the arrows (b).
The tail on the right hand side of the drift time spectrum originates from electrons drifting 
from the space between the detection planes.}
\label{dtime}
\end{figure}

\begin{figure}[ht]
\centering
\leavevmode
\epsfverbosetrue
\epsfclipon
\epsfxsize=6cm
\epsffile[5 5 500 480]{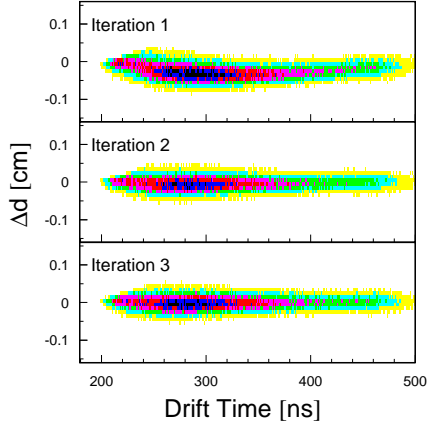}
\caption{Differences of the measured and fitted distances calculated as a function
of the drift time for the first detection plane and the angular bin 
$\theta \in (50^{\circ}, 60^{\circ})$ in three subsequent iterations of the calibration procedure.}
\label{correct}
\end{figure}

The chamber is calibrated using the experimental data.
In a first step an approximate drift time to drift distance relation is determined 
by integration of the drift time spectra as provided by the uniform irradiation method~\cite{Bs}.
A typical spectrum obtained by summing up drift time spectra from all sense wires
in one detection plane is shown in Fig.\ \ref{dtime}a.
The width of the drift time distribution is about 300~ns which with 
the corresponding half of the cell width equal to 9~mm results in a mean drift
velocity of 0.03~mm/ns.
However, the space-time relationship obtained by integration of the drift time spectrum is not
strictly linear (see Fig.\ \ref{dtime}b); 
its slope increases slightly in the neighborhood of the sense wire. 
In the next step, corrections to this calibration are determined using an iterative procedure. 
For this, the average deviations between the measured and fitted distances of the tracks from
the sense wires are calculated.
These deviations are determined as a function of the drift time $t_j$ corresponding to the TDC channel $j$
and the track entrance  angle $\theta_k$ from the range ($0^{\circ}$, $90^{\circ}$) 
divided into 9 intervals numerated by the index $k$:
\begin{equation}
\Delta d(t_j,\theta_k) = <\!d^f(t_j)\!> - d(t_j,\theta_k),
\end{equation}
where the average $<>$ is taken over all hits which were recorded in the TDC channel $j$ 
and correspond to the track entrance angle interval $k$.
The space-time relation used in the first iteration is then corrected by the above deviation:
\begin{equation}
d'(t_j,\theta_k) = d(t_j,\theta_k) +\Delta d(t_j,\theta_k) .
\end{equation}
After performing the reconstruction of tracks with the new space-time relation 
new corrections are calculated and so on. 
This procedure is repeated until the corrections become smaller than the position resolution
of the chamber. This occurs in our case after 2-3 iterations.
Fig.\ \ref{correct} shows differences $\Delta d$ of the measured and fitted distances calculated as a function
of the drift time for three subsequent iterations. The mean value of $\Delta d$ deviates from zero only after
the first iteration (upper panel in Fig.\ \ref{correct}) and the corresponding correction to the space-time relation is
of the order of 0.3~mm. For higher iterations the corrections are negligible. 
The standard deviation of $\Delta d$ is about 0.2~mm 
and is a measure of the single wire resolution. 

\begin{figure}[ht]
\centering
\leavevmode
\epsfverbosetrue
\epsfclipon
\epsfxsize=6cm
\epsffile[0 0 500 480]{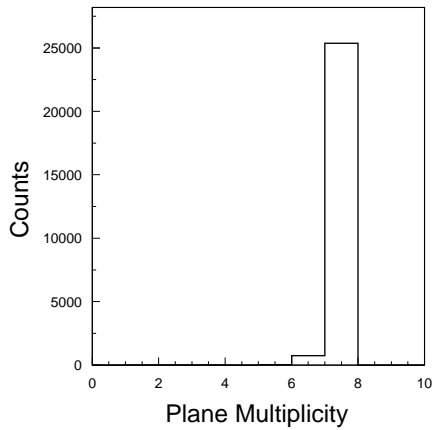}
\caption{Multiplicity of coincidences between the detection planes for single tracks 
of elastically scattered protons.}
\label{plmulti}
\end{figure}

\begin{figure}[ht]
\centering
\leavevmode
\epsfverbosetrue
\epsfclipon
\epsfxsize=6cm
\epsffile[55 170 520 640]{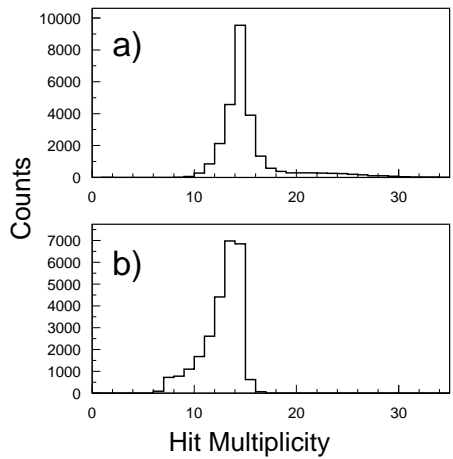}
\caption{Multiplicity of hits for events of the elastic proton-proton scattering (a) and of hits
preselected and used in successful reconstruction of single tracks (b).}
\label{hitmulti}
\end{figure}

For testing the track reconstruction we used single tracks of protons scattered elastically 
on the COSY-11 hydrogen target of the cluster-jet type at a beam momentum of 3.3~GeV/c. 
Events of the elastic proton--proton scattering
were registered as coincidences between the scintillation hodoscope S1
detecting the forward scattered protons and the monitor scintillator placed in 
the target region on the opposite
side of the beam, measuring the recoil protons. 
Additionally, we required firing of the seventh detection plane in order to assure that 
the proton tracks pass through the chamber. 
The detection efficiency per detection plane was estimated on the
basis of multiplicities of coincidences between the detection planes.
In 97\% of cases all seven planes fire (see Fig.\ \ref{plmulti}) which means that the detection efficiency
in a single detection plane is close to 100\%.
The multiplicity of hits shown in Fig.\ \ref{hitmulti}a is peaked around the value of 14 
which is due to the fact that most of tracks are inclined with respect to the chamber 
and mostly two neighbouring cells from one detection plane fire for each track.
For the reconstruction we used an ``effective'' position resolution of 220 $\mu$m
for which the mean value of $\chi^2$ per degree of freedom is in the vicinity of 1. 
The reconstruction was treated as successful when the $\chi^2$ per degree of freedom
was smaller than 5. 
This was the case for 95\% of events triggered as elastic $p-p$ scattering.
For selected events in which all seven detection planes fired, the reconstruction was successful in 97\%.
In the reconstruction most of the registered hits were used for the fitting as one can see by comparison of 
Fig.\ \ref{hitmulti}a and Fig.\ \ref{hitmulti}b.
The precision of the track parameters $a,b,c,d$ was extracted from the diagonal elements of the corresponding
covariance matrix. For typical track entrance angles of about $50^\circ - 60^\circ$ the resulting uncertainty 
of track position along the $x$- and $y$-axis is equal to $\delta b =$~0.3~mm and $\delta d =$~3~mm, respectively.
The uncertainty of the track inclination in the horizontal and vertical plane 
is about 1~mrad and 10~mrad, respectively.

\section{Conclusions}

A drift chamber with a c-shaped frame  was constructed for the COSY-11 experimental facility.
The special shape of the frame allows for the detection of low momentum particles without
disturbing the high momentum ejectiles which are registered in other detector components.
This kind of chamber can be applied if no frame elements are allowed
in the sensitive area of the detection system. 
An example of a possible further application --- besides the one discussed here --- 
could be the detection of particles scattered at small angles with respect to the beam. 
In this case, two such chambers placed symmetrically with respect to the beamline could 
be used for covering the forward angles. 
The chambers can be installed and removed without dismounting the beam-pipe
contrary to an alternative solution of one chamber with a central hole for the beam-pipe.

The track reconstruction algorithm which was developed for the chamber can also be used for other
planar drift chambers containing cells with an electric field which has approximately cylindrical 
symmetry. 
It is required, however, that the chamber contains wires oriented at least in three
different directions and for each direction there are at least two detection planes necessary.

The chamber allows to determine the track position and inclination in the horizontal plane with an
accuracy of about 0.3~mm and 1~mrad, respectively.
In the vertical direction these accuracies are worse by about an order of magnitude, 
which is in accordance with the design values.
For the tracking in the vertical magnetic field
of the COSY-11 dipole magnet no higher precision was envisaged. 
For other applications the precision can be improved by choosing a larger inclination of the wires.

{\bf Acknowledgments}

This work has been supported by the European Community - Access to
Research Infrastructure action of the
Improving Human Potential Programme,
by the DAAD Exchange Programme (PPP-Polen),
by the Polish State Committee for Scientific Research
(grants No. 2P03B07123 and PB1060/P03/2004/26)
and by the Research Centre J{\"u}lich.

\end{document}